\documentclass[usenatbib]{mn2e}
\usepackage{graphicx, amssymb, aas_macros, natbib, bm}
\newcommand{\mvir}{M_{\rm{vir}}}

\newcommand{\mstar}{M_{\star}}
\newcommand{\lstar}{L^{*}}
\newcommand{\msun}{{\rm M}_{\odot}}

\newcommand{\mpc}{{\rm Mpc}}
\newcommand{\kpc}{{\rm kpc}}
\newcommand{\kms}{{\rm km \, s}^{-1}}

\newcommand{\lcdm}{$\Lambda$CDM}
\newcommand{\diso}{d_{\rm iso}}
\newcommand{\niso}{n_{\rm iso}}

\title[A dichotomy in satellite quenching]
{
A Dichotomy in Satellite Quenching Around $\bm{\lstar}$ Galaxies
}
\author[Phillips et al.]
{John I.~Phillips,$^1$\thanks{$\!\!$e-mail: johnip@uci.edu} 
Coral Wheeler,$^1$ 
Michael Boylan-Kolchin,$^{1,2}$\thanks{$\!\!$centre for Galaxy Evolution Fellow} 
James S.~Bullock,$^1$ 
\newauthor Michael C.~Cooper,$^1$
 Erik J.~Tollerud$^3$\thanks{$\!\!$Hubble Fellow}\\
$\!\!^1$centre for Cosmology, Department of Physics and Astronomy, 
  4129 Reines Hall, University of California, Irvine, CA 92697, USA\\
$\!\!^2$Department of Astronomy, University of Maryland, College Park, MD 20742, USA\\
$\!\!^3$Astronomy Department, Yale University, P.O. Box 208101, New Haven, CT
06510, USA
}
\begin{document}

\pagerange{\pageref{firstpage}--\pageref{lastpage}} 
\pubyear{2013}

\maketitle

\label{firstpage}
\begin{abstract}
  
We examine the star formation properties of bright ($\sim 0.1 \, \lstar$)
satellites around isolated $\sim \lstar$ hosts in the local Universe using  spectroscopically confirmed systems in 
the Sloan Digital Sky Survey DR7.  Our selection method is carefully designed with the aid
of $N$-body simulations to avoid groups and clusters.
 We find that satellites are significantly more likely to be quenched than 
 a stellar mass-matched sample of isolated galaxies.
Remarkably, this quenching occurs only for satellites of hosts that
are themselves quenched: while star formation is unaffected in the
satellites of star-forming hosts, satellites around quiescent
hosts are more than twice as likely to be quenched than stellar-mass
matched field samples.  One implication of this is that whatever shuts down
star formation in isolated, passive $\lstar$ galaxies also plays at least an indirect
role in quenching star formation in their bright satellites.   The previously-reported tendency for
``galactic conformity'' in colour/morphology may be a by-product of this
host-specific quenching dichotomy.  
The S\'{e}rsic indices of quenched satellites
are statistically identical to those of field galaxies with the same
specific star formation rates, suggesting that environmental and
secular quenching give rise to the same morphological structure.
By studying the distribution of pairwise velocities between the hosts and satellites, we
find dynamical evidence that passive host galaxies reside in dark matter haloes
that are $\sim 45\%$ more massive than those of star-forming host galaxies of the same stellar mass.
We emphasize that even around passive hosts, the mere fact that galaxies become satellites
does not typically result in star formation quenching:  we find that only
$\sim 30 \%$ of $\sim 0.1 \lstar$ galaxies that fall in from the field are quenched around passive hosts, compared with $\sim 0\%$ around star forming hosts.  

\end{abstract}


\section{Introduction}
\label{sec:Intro}

In the now-favored dark energy plus cold dark matter (\lcdm)
cosmological model, the abundance of dark matter subhaloes as a
function of parent dark matter halo mass is predicted with high
precision in $N$-body simulations \citep[e.g.][]{bk09, klypin11}. In
contrast, the observed abundance and properties of satellite galaxies
 in the local Universe are
poorly understood within the current cosmological framework. For
example, modern semi-analytic \citep[e.g.][]{somerville08, guo11} and
hydrodynamic \citep[e.g.][]{dave11} models of galaxy formation are
overly effective at halting star formation in satellite systems, such
that current models dramatically overpredict the number density of
quenched (or passive) satellites at $z \sim 0$ \citep{kimm09,
  weinmann06, weinmann10, weinmann12}.

Additionally, current models of galaxy formation fail to reproduce the
relationship between host and satellite galaxy properties. In
particular, recent studies of satellite and central galaxies in the
Sloan Digital Sky Survey \citep[SDSS,][]{york00} find that satellites
of red (or passive) host galaxies are more likely to also be passive
relative to their counterparts around star-forming hosts
\citep{weinmann06}. This correlation between host and satellite
properties, commonly dubbed ``galactic conformity,'' is also poorly
replicated by modern semi-analytic models \citep{kauffmann13},
potentially due to a lack of decoupling between the growth of
satellite galaxies and the growth of their dark matter haloes in the
models \citep{weinmann12}. Ultimately, understanding this 
connection between the properties of host and satellite galaxies may
be a powerful way to constrain the physical mechanisms responsible for
halting star formation in passive systems.

To better understand the physics driving the relationship between the
properties of host and satellite galaxies, it is particularly
interesting to examine the satellite populations around $\lstar$
hosts. At this luminosity (or stellar mass), central galaxies are
roughly equally likely to be star forming or quenched, while at higher
(or lower) luminosities, the host population is increasingly dominated
by passive (or star forming) systems
\citep[e.g.][]{kauffmann03, bell03, baldry04}. As such, this scale
($\sim \! \lstar$) is well-suited for exploring connections between host
and satellite properties. Moreover, our own Milky Way falls into this
class of galaxies; understanding the statistical properties of
$\lstar$ galaxies will therefore inform our understanding of the Milky
Way and its satellites, the objects that can be studied in the most
detail.

In this work, we investigate satellite quenching mechanisms by
comparing satellites around isolated $\lstar$ galaxies to a statistically identical sample of field galaxies. 
In doing so, we build upon the work presented in \citet{tollerud11b},
who used strict isolation criteria to explore satellite galaxy counts 
around galaxies in the field.
By focusing on
isolated hosts, we are able to eliminate the known effects of massive
haloes (groups or clusters) and large-scale structure on satellites. We further split the
samples based on the star forming activity of the host to investigate
the origins and strength of correlations between star formation
properties of satellites and hosts (i.e.~the physical drivers of
galactic conformity). In Section 2, we introduce our observational 
sample, use N-body simulations to develop criteria for
selecting isolated $\lstar$ hosts, and quantify the expected level of
interloper contamination.   In Section
\ref{sec:iso_sdss}, we describe the selection criteria as applied to
the SDSS to create our galaxy samples. In Section \ref{sec:results},
we present our findings on the relationship between star formation
properties of satellites and their host galaxies. Finally, we discuss and
summarize the implications of the observed trends on galaxy formation
models in Section
\ref{sec:discuss} and Section \ref{sec:summary}. Unless otherwise noted, all logarithms are base
10. Halo virial masses are defined with respect to an overdensity of 94 relative to the critical density of the universe \citep{bryan98}. Throughout our analysis, we employ a $\Lambda$ cold dark matter ($\Lambda$CDM)
cosmology with WMAP7+BAO+H0 parameters  $\Omega_{\Lambda} = 0.73$, $\Omega_{m} =
0.27$, and $h =0.70$ \citep{BAO}.

\section{Using Simulations to Define the Observational Sample}
\label{section:sim}

Before we detail our approach to isolate host galaxies and their
satellites, a brief description of our adopted terminology is in order.

We use the term ``satellite" to refer to a galaxy that is within the virial volume of
a central ``host" galaxy's halo.  Given that the observational
data are subject to projection effects, we refer to our raw
sample of observed
galaxy pairs as ``primaries" and ``secondaries".  
We also construct theoretical mock catalogs, and use 
the terms host and satellite once the
full three-dimensional information in the simulation is used to confirm whether 
an ``observed"  secondary is actually within the virial volume of the primary.
Finally, based on these mock observations, our
observationally-derived statistics will be corrected for contamination effects;
accordingly, we will use the terms ``host'' and ``satellite'' in
discussing contamination-corrected observed parameter distributions as well.

\subsection{Observational Data}

In selecting our observational sample, we employ data from Data
Release 7 (DR7) of the Sloan Digital Sky Survey \citep[SDSS][]{york00,
  abazajian09}. In particular, we utilize the MPA-JHU derived data
products, including median total stellar masses, photometrically
derived according to \citet[][see also
\citealt{salim07}]{kauffmann03}, and median total star formation
rates, measured from the SDSS spectra as detailed by
\citet{brinchmann04}. Note that their procedure includes methods for estimating specific star formation rates when they are unobtainable from emission lines. Supplemental information is drawn from the NYU
Value-Added Galaxy Catalog \citep[NYU-VAGC][]{blanton05vagc}, such as
S\'{e}rsic indices \citep{sersic68} that are derived from
one-component fits according to \citet{blanton03, blanton05}.

As a first step in identifying satellite and host galaxies, we compile a list of  secondaries containing
all galaxies with a stellar mass of $10^{9.5} \msun< \mstar <
10^{10.5} \msun$ and a list of primaries with $\mstar >
10^{10.5} \msun$. As discussed in more detail in Section
\ref{subsec:isolated}, we then apply isolation criteria to the
potential primary sample and spectroscopically search for
physically-associated secondaries. We also impose a spectroscopic completeness (\emph{fgotmain}) cut of  0.7. This cut corresponds to a mean sample spectroscopic completness of 92\%, making it extremely unlikely that an object for which no redshift was obtained would impact our isolation procedure. Furthermore, we do not expect the small remaining incompleteness to bias our satellite selection. Our final samples of primary and
secondary galaxies have mean stellar masses of $\mstar =
10^{10.80} \msun$ and $ \mstar = 10^{9.97} \msun$, respectively. We impose a
limiting redshift of $z = 0.032$, within which we are complete to a
stellar mass of $10^{9.5}~\msun$. The mean redshift of our sample is
$0.027$. Our final sample contains $457$ primary/secondary
systems. These parameters are similar to the selection
criteria described in \citet{tollerud11b}.

We further divide the primaries into active and passive categories,
with the dividing line between the the two star formation rate classes given by
\begin{equation} 
\label{eq:thresh}
\log({\rm SSFR}_{\rm host}) = -0.6\, \log(M_{\star, \rm host}) - 5.2, 
\end{equation}
where SSFR$_{\rm host}$ denotes the specific star formation rate of
the host galaxy. This equation is motivated by the established blue
cloud/red sequence dichotomy of galaxies in the SDSS
\citep{strateva01, baldry04, blanton05}. The slope was selected to
match the slope of the red sequence in our sample.% (see Figure
%\ref{fig:samp}).
% \textbf{mbk: short description of motivation for this equation}.

\subsection{Numerical Simulation}
\label{sec:num_sim}
Our goal of studying satellites of isolated $\lstar$ galaxies requires
a rigorous and accurate identification of truly isolated $\lstar$
primaries, systems that are largely excluded from residing within group and cluster
environments.   Accordingly, we use the Millennium-II Simulation
(hereafter, MS-II; \citealt{bk09}) to test our selection and isolation
procedures. The MS-II has a box size of $137$ Mpc, 
well-matched to our observational volume at $z < 0.032$. Its high mass
resolution -- $m_p=9.4\times 10^{6}\,\msun$ -- ensures that it is
complete for halo masses in excess of $\sim 2\times 10^{10}\,\msun$,
corresponding to stellar masses of $\sim 10^{7.5-7.75}\,\msun$
\citep{guo10, behroozi10, behroozi12, leauthaud12, moster13}. Using
the MS-II, \citet{tollerud11b} demonstrate the effectiveness of this
technique in obtaining clean samples of isolated hosts with halo
masses of roughly a few $\times 10^{12} \, \msun$.

Following the abundance matching prescription of \cite{guo10}, we find
that our lower limit for the stellar mass of secondaries ($\mstar =
10^{9.5} \msun$) corresponds to $v_{\rm max} = 94.8\, \kms$. The
cutoff between what we consider to be a primary and a secondary,
$v_{\rm max,cut}$, % is set as described in Section \ref{sec:iso} and
corresponds to a value of $v_{\rm max,cut} = 166.5\, {\kms}$.
We use these values to construct mock catalogs below.

%%%%%%%%%%%%%%%%%%%%%%%%%%%%%%%%%%%%%%%%%%%%%%%%%%%%%%%%%%%%%%
\begin{figure}
 \centering
 \includegraphics[scale=0.50, viewport=20 0 800 410]{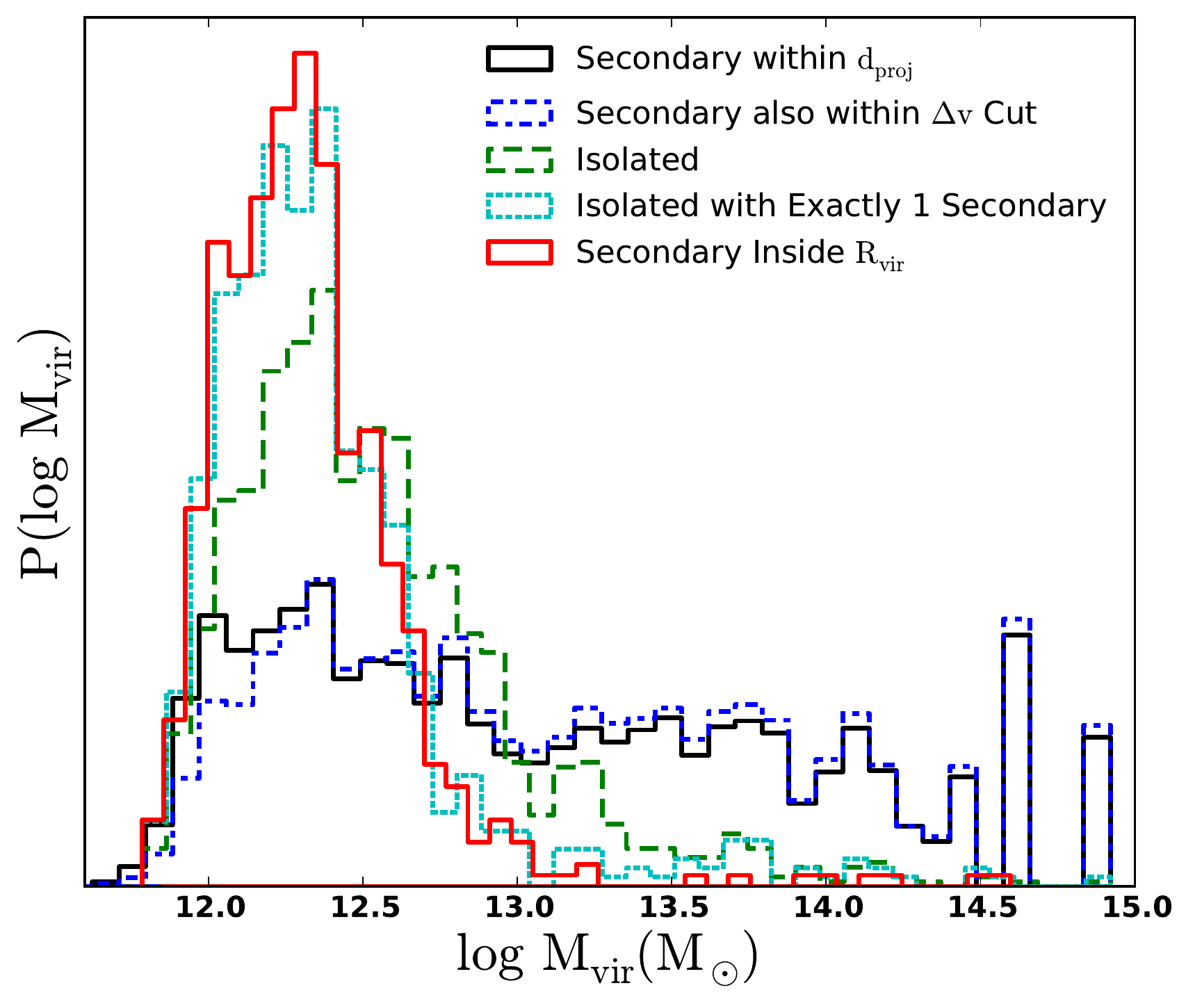}
 \caption{
 Probability distribution of virial masses of primaries
   associated with each secondary within a projected distance of
   $d_{\rm proj} = 355\, {\kpc}$ for various selection cuts in the
   MS-II mock catalogs. Using only this $d_{\rm proj}$ criterion (black solid line),
   many of the selected secondaries are associated with very massive dark matter
   host primaries. Adding a restriction of $\Delta v_{\rm sel} = 500\,\kms$
   does not help in isolating MW-size hosts, as cluster and group haloes
    typically contain galaxy pairs that satisfy this criterion.  Further
   restricting this subset of primaries to those that that are
   isolated (according to our fiducial set of isolation criteria as
   defined in Section~\ref{subsec:purity}) \textit{significantly}
   reduces the high-mass tail (green dashed line).  Adding the
   constraint that primaries have \textit{exactly} one secondary (cyan
   dotted line) further decreases the number of group-mass ($\sim \!
   10^{13}\,\msun$) haloes selected. This line represents the halo mass distribution for our full sample. The red line demonstrates that in
   most group-scale systems in our sample, the secondary is outside
   the virial radius of the group's central halo. Thus, even for the
   small number of group-sized haloes in our sample ($< 7\%$ of secondaries), we do not expect
   physical mechanisms specific to the group environment to
   significantly bias our results.
 \label{fig:mtop}
}
\end{figure}
%%%%%%%%%%%%%%%%%%%%%%%%%%%%%%%%%%%%%%%%%%%%%%%%%%%%%%%%%%%%%% 

\subsection{Obtaining Isolated $\bm{\lstar}$ Galaxies}
\label{subsec:isolated}

Figure~\ref{fig:mtop} illustrates the difficulty in selecting satellite galaxies around isolated $\lstar$ host
haloes without significant contamination by cluster
or group members.    We use the MS-II simulation to perform a series of mock observations
to identify secondaries and then determine the host virial
mass of the associated subhaloes (horizontal axis) for various
selection criteria.    The black solid histogram shows the host halo virial mass distribution for all
secondaries  within a projected distance of $d_{\rm
  proj} = 355\, {\kpc}$ of an identified primary.  We count one host halo for each secondary identified, implying that that same host halo can appear more than once if it hosts multiple secondary candidates.  
 It is evident that this selection picks out a large
percentage of satellites that reside within group and cluster-mass haloes $(\mvir >
10^{13}\,\msun)$. 

A standard additional criterion for spectroscopic secondary selection is a cut on the velocity difference between the primary
and secondary, $\Delta v$. As we might expect satellites of group sized haloes to be kinematically warm, this cut could prove useful for excluding such
massive hosts. However, restricting pairs to lie within a projected velocity limit $\Delta v_{\rm sel} =
500\,\kms$ only weakly eliminates haloes within the cluster and group
mass regime (blue dashed histogram), while preferentially removing galaxy-sized $(< \!
10^{12.5}\,\msun)$ haloes (see blue dashed histogram in
Fig.~\ref{fig:mtop}). Such a cut on velocity difference is ineffective
at removing higher-mass haloes since group environments contain many
objects -- with a wide range of velocities -- in the secondary range.
As such, it is very likely that one of these objects will meet both
our projected distance and projected velocity criteria, even if many
or even most will not.

In order to filter out satellites associated with massive host haloes, we adopt an approach pioneered
by \citet{Barton07}, which relies on strict isolation cuts \citep[see also][]{tollerud11b,Edman12}.
Specifically, we impose limitations on $\niso$, the maximum number of other haloes (galaxies) in the
primary $v_{\rm max}$ ($M_\star$) range that we allow within an annulus bounded on
the interior by $d_{\rm sel}$ and on the exterior by $\diso$, and bounded in velocity space by $\pm \Delta v_{iso}$. We never
allow any other primaries within $d_{\rm sel}$ in projected distance space and $\pm \Delta v_{iso}$ in velocity space.  The green dashed
histogram in Figure~\ref{fig:mtop} shows the host halo mass
distribution when we restrict our selection to only those haloes that
(1) have no other primaries within $d_{\rm sel} = 355\, {\kpc}$ and
$\Delta v_{\rm iso} = 1000\, {\kms}$ and (2) have no more than one
other primary within $d_{\rm iso} =1\, \mpc$ and $ \Delta v_{iso} =
1000\,{\kms}$. We will show below that this particular set of criteria
are optimal for isolating MW-sized haloes and their LMC-sized companions, and also for maximizing the purity of our pair
sample. With the isolation criteria imposed, we greatly reduce the
number of haloes in our sample that inhabit clusters and
 somewhat reduce those in the group mass regime as well.

As a final cut, we select only those haloes that have \textit{exactly}
one secondary within $355\, \kpc$ and $|\Delta v| <
500, \kms$. Applying this additional selection criterion (see the
cyan dotted histogram in Fig.~\ref{fig:mtop}) removes nearly all pairs
that fall within groups and cluster, yielding a population dominated
by Milky Way-like haloes; only 6.7\% of objects satisfying our
isolation and satellite number cuts occupy haloes with $\mvir >
10^{13}\,\msun$ within the MS-II. Moreover, for the small tail of secondaries associated with massive primaries, only 38\% of them actually 
lie inside the virial radius of the identified primary -- the rest sit at the
outskirts of their associated friends-of-friends group in the MS-II catalog.
The red solid histogram in Figure \ref{fig:mtop} shows the mass distribution
for secondary hosts that lie within the virial radius of their
respective host haloes.  The resultant histogram clearly picks out the galaxy-mass
scale for hosts. The median host mass for this distribution is $\rm M_{\rm vir} = 
10^{12.29} \, \msun$ with a 68\% spread of $\rm 10^{12.17} - 10^{12.40} \, \msun$. While it would be impossible to employ this in our observational sample, we do not expect satellites that lie outside of their host's virial radius to contribute significantly to a quenching signal \citep{Wetzel13b}.

The selection cuts illustrated by the cyan histogram in Figure~\ref{fig:mtop} represent our fiducial choices
for five parameters that we have tuned to identify exclusively 
bright satellites around isolated $\sim \lstar$ hosts.  These choices
are based on tests against mock observations in the MS-II simulation:
\begin{itemize}
\item $d_{\rm sel}=355\,\kpc$: The maximum projected
  separation between primary and secondary.
\item $\diso = 1\,\mpc$: The maximum projected distance within which
  we check for neighbouring primaries.
\item $\niso=1$: The maximum number of neighbouring primaries allowed between $d_{\rm
    proj}$ and $\diso$ within $\Delta v_{\rm iso}$ in velocity. We always require that no neighbouring primary can fall
  within $d_{\rm sel}$ in projected distance and $\Delta v_{\rm iso}$ in velocity, such that the primary in our sample
  have no other primary within a distance of $d_{\rm sel}$.
\item $\Delta v_{\rm sel}=500\,\kms$: The maximum line-of-sight
  velocity difference between primary and secondary.
\item $\Delta v_{\rm iso}=1000\,\kms$: The minimum line-of-sight
  velocity difference required between the primary and any other neighbouring primary
  within a projected distance of $\diso$ to not count towards $\niso$. No neighbouring primary is allowed within $d_{\rm sel}$ and $\Delta v_{\rm iso}$.
\end{itemize}

In the next subsection we illustrate how these fiducial choices were motivated
to balance purity and sample size.

%%%%%%%%%%%%%%%%%%%%%%%%%%%%%%%%%%%%%%%%%%%%%%%%%%%%%%%%%%%%%%
\begin{figure}
 \centering
 \includegraphics[scale=0.41, viewport=20 0 800 410]
 {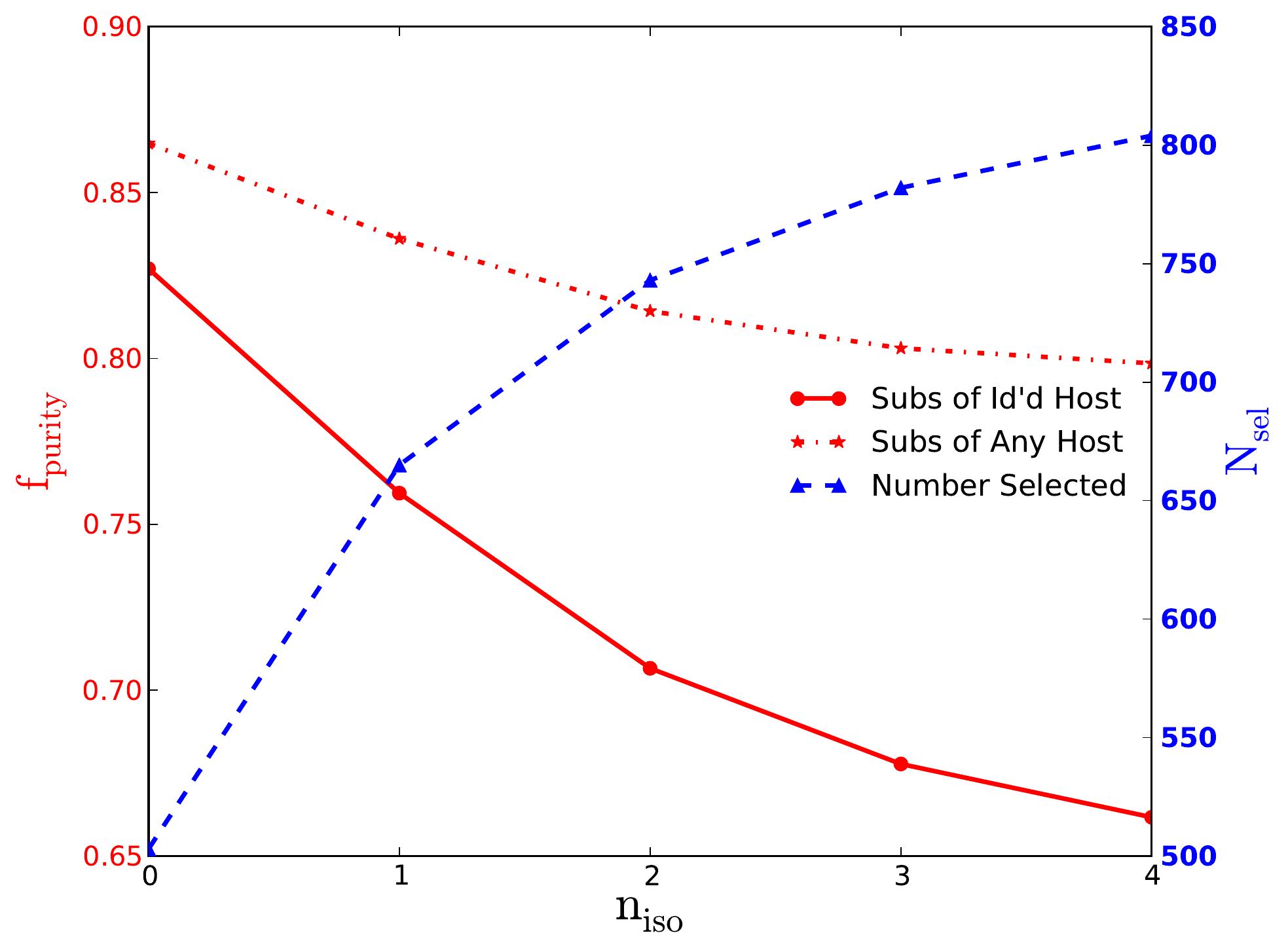}
 \caption{Fraction of the pair sample (in the MS-II) for which the secondary is a
   subhalo of the primary (red solid line) and for which the secondary
   is a subhalo of any host (red dash-dotted line) alongside the
   number of pairs selected (blue dashed line) as a function of one of our isolation criteria, $\niso$. While purity rises
   sharply as $\niso$ approaches $0$, the number of pairs selected
   decreases precipitously. As a compromise between sample size and
   purity, we define our selection limit to be $\niso =
   1$. At this value of $\niso$,  the secondary is a subhalo of the
   primary in $75\%$ of pairs while the secondary is a subhalo of
   any host in just over $85\%$ of the sample.
 \label{fig:nout}
}
\end{figure}
%%%%%%%%%%%%%%%%%%%%%%%%%%%%%%%%%%%%%%%%%%%%%%%%%%%%%%%%%%%%%% 

\subsection{Balancing Purity and Sample Size}
\label{subsec:purity}
A crucial characteristic of any observational sample is its purity, $f_{\rm
  purity}$, which we define within the MS-II to be the fraction of
identified host-satellite pairs having (1) primaries that are actually host haloes (i.e.~not satellites of a
larger system); (2) secondaries that are actually satellites (i.e.~not
lower-mass primaries); and (3) satellites that are actually satellites
of the host (and not satellites of a different host that are included
because of projection).
All of the parameters listed in Section \ref{subsec:isolated} could be
tuned to increase the purity of the sample, however this usually
involves a trade-off with regard to sample size. As an illustration,
we present one such example of this tuning: Figure \ref{fig:nout}
illustrates the dependence of sample purity on our isolation
criterion, $\niso$, holding all other parameters fixed at their
fiducial values. As expected, increasing $\niso$ increases the number
of interlopers and decreases $f_{\rm purity}$ (solid red line). While
the purity of our sample increases with the strictness of our
isolation criterion, $\niso$, the sample size simultaneously decreases
(blue dashed line). As a compromise between sample size and purity, we
define our selection limit to be $\niso \leq 1$ throughout the
remainder of this work, yielding a sample purity of 75\%. While using
$\niso=0$ would increase the purity to 82\%, it would also reduce the
sample size by more than 25\%.

Note that if we relax the definition of a true pair to require only
that the secondary be a subhalo, rather than being a subhalo of the
chosen host, our purity reaches approximately 85\% and depends little
on $\niso$ (dotted red line). In subsequent sections, we perform
corrections to account for the presence of interlopers in the
observational data using this modified definition of purity. We 
make the assumption that the interloper population is comprised only
of field galaxies having the same specific star formation rate
distribution as the control sample (see Section~\ref{sec:results}).

%%%%%%%%%%%%%%%%%%%%%%%%%%%%%%%%%%%%%%%%%%%%%%%%%%%%%%%%%%%%%%
\begin{figure}
 \centering
 \includegraphics[scale=0.42, viewport=20 0 800 410]{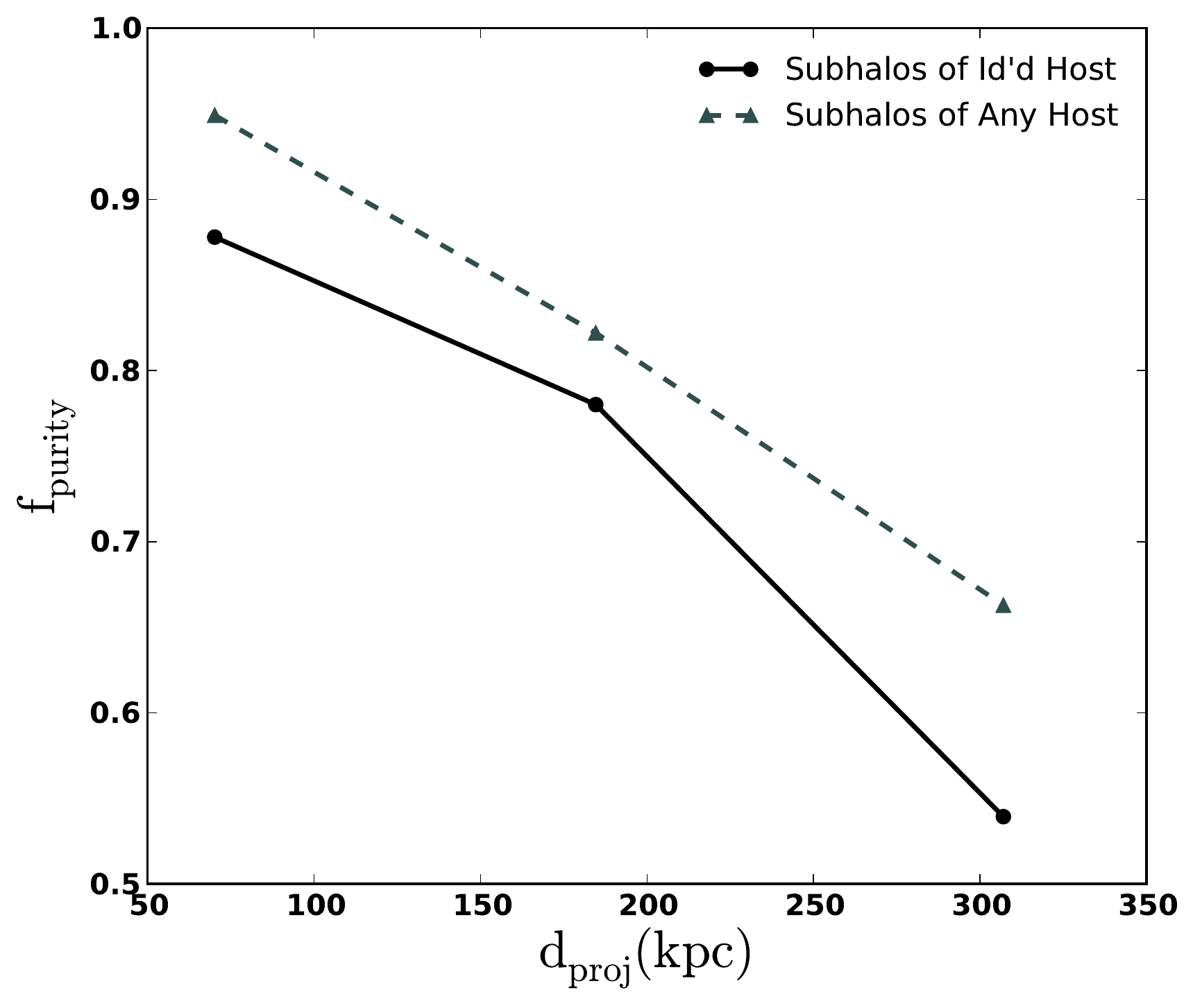}
 \caption{Fraction of the pair sample for which the secondary is a
   subhalo of the primary (solid line) and for which the
   secondary is a subhalo of any host (dashed line) as a function
   of projected distance between the primary and secondary in the
   MS-II. Due to an increased chance of interlopers at larger $\rm
   d_{proj}$, the purity 
   decreases with projected distance from the primary. The fraction of
   secondaries that are true subhaloes of the primary decreases from $\sim
   \! 88\%$ in the innermost bin to just above $\rm 50\%$ at $\rm
   355\, {\kpc}$, while the fraction of secondaries that are subhaloes
   of any host decreases from $\sim \! 96\%$ in the innermost bin to
   $70\%$ at $\rm 355\, {\kpc}$. 
 \label{fig:projd}
}
\end{figure}
%%%%%%%%%%%%%%%%%%%%%%%%%%%%%%%%%%%%%%%%%%%%%%%%%%%%%%%%%%%%%% 

Purity could also vary within our sample as a function of separation
between the primary and secondary galaxy, either in physical space or
velocity space. While sample size and purity exhibit little dependence
on $\Delta v$, we find significant variation in purity with the
projected separation between the primary and secondary. As shown in
Figure \ref{fig:projd}, sample purity increases with decreasing
$d_{\rm proj}$ for both our standard and less-restrictive definition
of purity (blue solid and red dashed lines, respectively). For close
separations ($\sim 60 \,\kpc$), approximately 88\% of secondaries are
subhaloes of their respective primaries. At larger distances, however,
the problem of contamination increases noticeably. At $\sim \rm 200\,
{\kpc}$, the purity drops to $\rm \sim 77\%$, while it falls to just
over 50\% at $300\,\kpc$. Even when considering whether the secondary
is a subhalo of any host, the purity decreases from $\sim 96\% $ in
the innermost bin to $86\%$ at $200\, \kpc$ and 70\% at $300
\,\kpc$. When we consider trends in quenching with projected radius,
we will take this radial dependence on purity into consideration.

%%%%%%%%%%%%%%%%%%%%%%%%%%%%%%%%%%%%%%%%%%%%%%%%%%%%%%%%%%%%%%
\begin{figure*}
 \begin{centering}
 \includegraphics[scale=0.44]{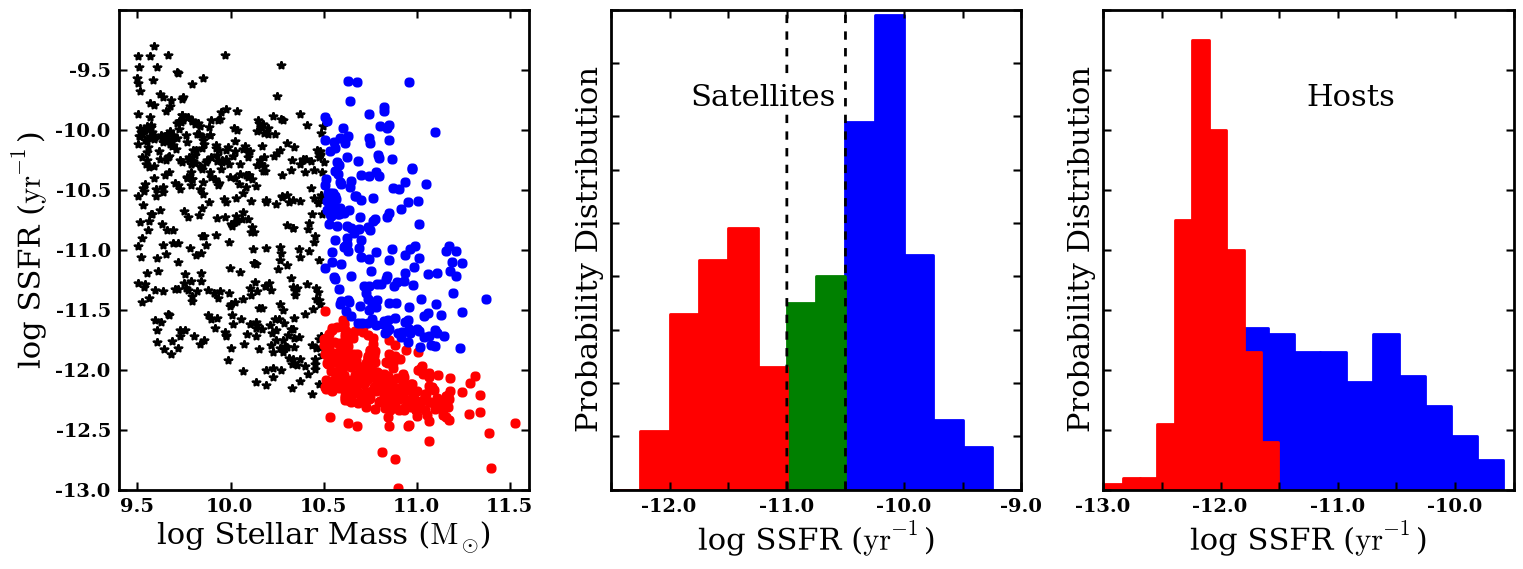}
 \end{centering}
 \caption{
 (\emph{Left}): The distribution of specific star formation
   rate (SSFR) as a function of stellar mass for the galaxy sample
   employed in this study. Galaxies with $\mstar > 10^{10.5} \msun$
   are categorized as hosts (red and blue points), with galaxies at
   lower mass categorized as satellites (black stars). Hosts are
   divided into two samples, where those coloured blue are star forming
   and those coloured red are passive. Each host has exactly one
   satellite. (\emph{Centre}): The SSFR distribution for
   the satellite population. For the purposes of quantifying quenching
   effects, we subdivide the satellite population into three categories of
   star formation activity: satellites with SSFRs greater than
   $10^{-10.5}~{\rm yr}^{-1} $ are defined as vigorously star forming
   (blue histogram), satellites with SSFRs between $\rm 10^{-10.5}$
   ${\rm yr}^{-1} $ and $10^{-11}~{\rm yr}^{-1} $ are defined as
   moderately star forming (green histogram), and satellites with
   SSFRs less than $\rm 10^{-11}~{\rm yr}^{-1}$ are deemed passive
   (red histogram). (\emph{Right}): Distribution of SSFRs for host
   galaxies. Hosts are divided into passive (red histogram) and
   star forming (blue histogram) subsamples according to blue
   cloud/red sequence membership as detailed in Equation 1. 
 \label{fig:samp}
}
\end{figure*}
%%%%%%%%%%%%%%%%%%%%%%%%%%%%%%%%%%%%%%%%%%%%%%%%%%%%%%%%%%%%%% 

\section{Isolated $\bm{\lstar}$ hosts in SDSS}
\label{sec:iso_sdss}
\subsection{Isolation Procedure Applied to Data}

Having identified an optimal set of selection criteria, we turn to applying these criteria to our observational data. To identify a set of potential primary galaxies, we begin by selecting
all galaxies in the SDSS DR7 catalog with stellar mass greater than
$10^{10.5} \msun$. The isolation criteria described in
Section~\ref{subsec:isolated} are then applied. In particular, we
discard (1) all primaries with one (or more) neighbour of stellar mass
$\mstar > 10^{10.5} \msun$ within a projected distance of $355\,\kpc$
and within a velocity difference of $|\Delta v| \equiv c |z_{1} -
z_{2}|<1000\, {\kms}$ along the line-of-sight, and (2) all primaries
with two (or more) neighbours of stellar mass $10^{10.5} \msun$ having $ 355\, \kpc < d_{\rm proj} < 1\, \mpc$ and $|\Delta v| <
1000\, \kms$. All galaxies that pass the isolation criteria are deemed
``isolated host-mass primaries.'' We then compile a catalog of
host/satellite pairs, our ``pairs catalog,'' by selecting those
isolated host mass primaries with exactly one neighbour in the stellar
mass range $10^{9.5} \msun < \mstar < 10^{10.5} \msun $ with $d_{\rm proj} <
355\,\kpc$ and $|\Delta v| < 500\, \kms$.  As shown in the previous section,
the requirement of exactly
one satellite allows us to better select hosts that have MW-sized
haloes (i.e.~$\mstar \sim 10^{12} \msun$). After the full isolation
procedure, our pairs catalog consists of 483 host-satellite
partners. Based on the results of Section~\ref{subsec:isolated}, we
expect that the hosts have a median virial mass of 1.9 
$\times 10^{12}\,\msun$.

\subsection{Observational Sample}
\label{sec:obs_sample}

The left panel of Figure~\ref{fig:samp} shows our hosts and satellites
in the specific star formation rate-stellar mass plane. Black stars
denote the satellite population, while our host population is coloured
by star formation properties (see Eqn.~\ref{eq:thresh}): red and blue
circles correspond to the passive and star forming hosts,
respectively. Hosts with multiple bright satellites and satellites
near two host-like objects were removed from the sample, such that
there is a one-to-one correspondance between the hosts and satellites
in the pairs sample.

The distribution of specific star formation rates for the satellite
population is given in the central panel of Figure~\ref{fig:samp}. We
place vertical lines at $10^{-11}$ $\rm yr^{-1}$ and $10^{-10.5}$ $\rm
yr^{-1}$ in this and all subsequent plots as the borders of our
defined regimes of star formation activity for the satellite
population: passive (red histogram), moderately star forming (green
histogram), and vigorously star forming (blue histogram).  Finally, the SSFR distribution for the hosts
is shown in the right panel of Figure~\ref{fig:samp}. Hosts are
divided into passive and star forming categories according to
Equation~\ref{eq:thresh}. 

\subsection{Satellite-host velocity distributions}

\begin{figure}
 \centering
%, viewport=20 0 400 420
  \includegraphics[scale=0.50, viewport=20 0 800 410]{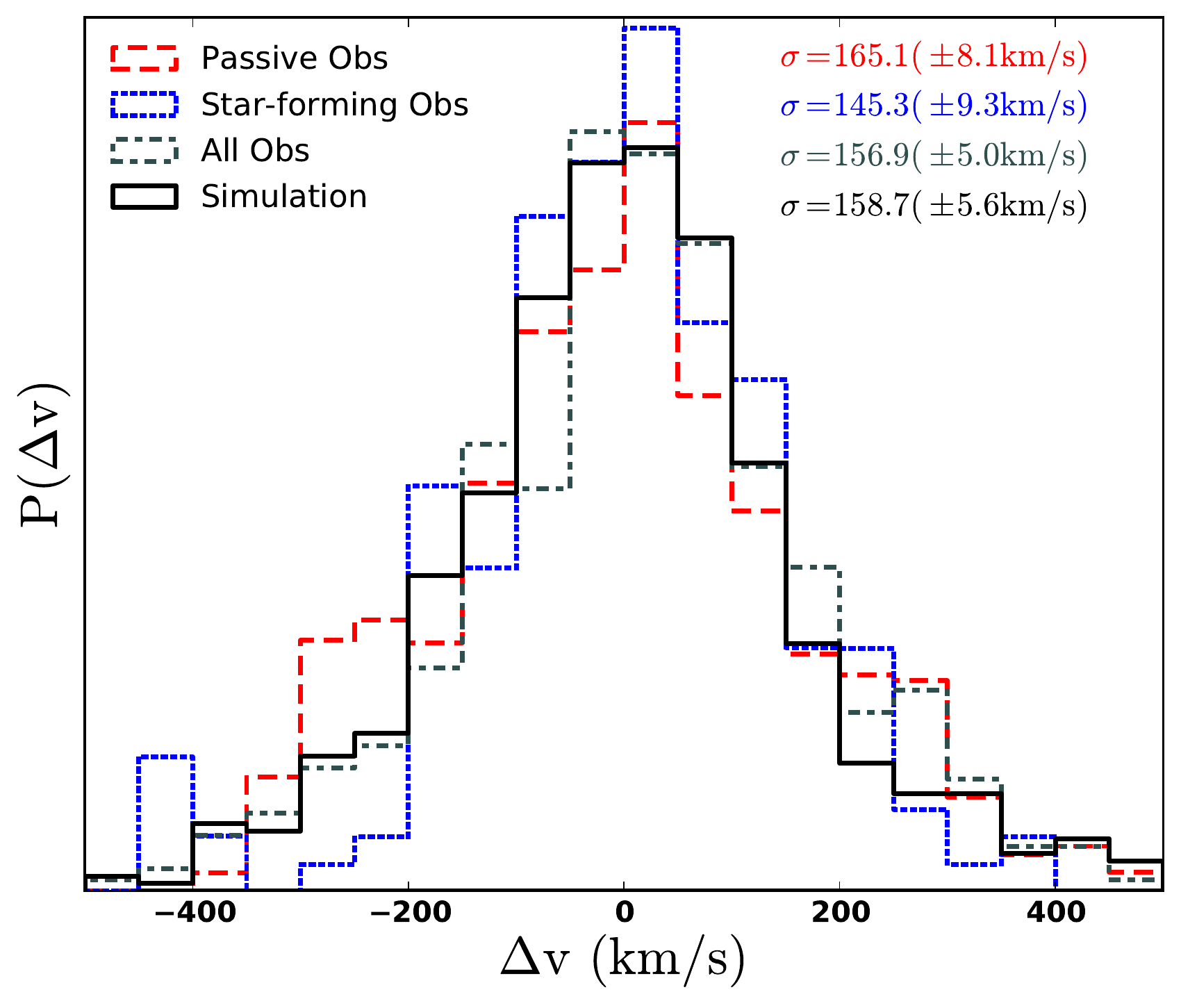}
 \caption{Probability distribution of velocity offsets $\Delta v$ for all observed primary-secondary pairs in our sample (dash-dotted grey) along with pairs chosen from stellar mass-matched samples of star forming (blue) and passive (red) hosts.  The root-mean-square of each distribution is given in the figure.  The passive host subsample appears kinematically warmer than the star forming host subsample, suggesting that these passive galaxies reside in more massive dark matter haloes at fixed stellar mass. Also shown are the host-satellite velocity distributions for our mock catalogs with the same selection criteria applied to MS-II (black line), which matches well the observed sample.  This remarkably consistency shows that our
 selection procedure is identifying host haloes as expected, with a virial mass distribution consistent with that shown by the red histogram in Figure 1.
  \label{fig:satellite_velocities}
}
\end{figure}

In Figure \ref{fig:satellite_velocities}, we show the distributions of $\Delta v$ for our full primary/secondary sample (dash-dotted grey) along with those for our star forming primary sample (blue curve) and a stellar mass-matched subsample of our passive primaries (red curve).  These
histograms represent ``stacked" satellite velocity distributions for hosts.  By bootstrapping our distributions, we find evidence (at approximately two standard deviations) that passive hosts reside in slightly more massive haloes {\em at fixed stellar mass}, with measured velocity dispersions $\sigma_{\rm passive} = 165.1 \pm 8.3$ km/s and $\sigma_{\rm star forming} = 145.4 \pm 9.4$ km/s.  This suggests that the haloes of passive $\lstar$ galaxies are $\sim 45 \%$ more massive at fixed stellar mass than those of star forming galaxies, assuming a mass-scaling proportional to $\sigma^3$, where the exponent is derived from  the assumption of a linear relation between halo maximum circular velocity and the spread of stacked pairwise host/satellite velocity distributions at fixed halo mass. \footnote{The qualitative result that passive hosts reside in kinematically warmer and thus more massive haloes is robust to the exact nature of the scaling between stacked satellite velocities and halo mass. We have used simulated observations described in Section \ref{sec:num_sim} to compare the observed velocity dispersions of satellite subhalos within hosts stacked in virial mass bins and found that the naive scaling adopted here does hold for the halo mass range of interest.}
 Also plotted is the distribution of velocity differences for the objects in MS-II chosen by mock observations
 that mirror our selection criteria (black histogram). Remarkably, the mock catalog falls almost perfectly between the red and blue histograms, providing an independent dynamical verification that we are indeed selecting the haloes we have set out to study using abundance matching.

The evidence we see for a halo mass trend with SSFR at fixed stellar mass is
qualitatively consistent with the results of \citet{mandelbaum06}, who find that early-type galaxies with stellar
masses of $\sim 10^{11}~\msun$ tend to live in haloes that are $\sim 3$ times more
massive than their star forming counterparts at the same stellar
mass.    The \citet{mandelbaum06} work samples all environments and thus the mass trend may 
be a result of the  colour-density relation where red galaxies prefer overdense regions -- i.e.~more massive haloes \citep{hogg04, cooper06, cooper10b, cooper10}. Our sample focuses on lower-mass systems, and specifically
avoids group and cluster haloes.

\subsection{Control Sample}
\label{section:control}

As a reference sample by which to compare our satellite population, we
define a set of galaxies in the field with stellar masses similar to
that of our satellite galaxies. As with our pairs sample, we tune our
selection criteria by maximizing purity, where purity here is defined to be the
fraction of objects that are actually hosts (i.e.~centrals) and not
satellites within the MS-II. Since we have no shortage of objects
(SDSS contains $12447$ such objects in our stellar mass range),
conserving sample size is less important, and we can tune our
selection parameters to create a maximally pure control sample. The final selection
parameters for the control sample are $n_{\rm iso, control} = 0$,
$d_{\rm iso, control} = 2.9\,\mpc$ and $ \Delta v_{\rm iso,
  control} < 400\, {\kms}$. With these values for the selection and
isolation criteria, we obtain $ f_{\rm p, control} = 0.97$. The
impurity in the sample is approximately equal to the Poisson noise
from the number of objects, $1 - f_{\rm purity, control}=1/\sqrt{N}$.

%%%%%%%%%%%%%%%%%%%%%%%%%%%%%%%%%%%%%%%%%%%%%%%%%%%%%%%%%%%%%%
%%%%%%%%%%%%%%%%%%%%%%%%%%%%%%%%%%%%%%%%%%%%%%%%%%%%%%%%%%%%%% 

Implementing these selection criteria, we construct a catalog of
isolated galaxies in the mass range $10^{9.5} < \mstar < 10^{10.5}$
to use as our control sample. Our isolation criteria are similar to,
but more stringent than, those used in selecting the pairs catalog. We
reject any galaxy with a neighbour of stellar mass $\log (\mstar /
\msun)> 9.5$ with $d_{\rm proj} < 2.9~{\mpc}$ and with $ \Delta v_{\rm iso} <
400\, {\kms}$. This catalog, which we will call the ``field dwarf
catalog" contains 581 galaxies, with a mean stellar mass of
$10^{9.95}~\msun$ and a mean redshift of 0.024.

\subsection{Parameter-Matching Procedure}

It will frequently be useful in our analysis to explore the variation
in a given galaxy property (e.g.~specific star formation rate) across
two related samples of galaxies (e.g.~isolated galaxies versus
satellites). When performing such a comparison, we would like to
control for correlations between the varied parameter and other
properties of the chosen samples --- for example, the observed
decrease in SSFR with increasing stellar mass \citep{brinchmann04,
  noeske07, elbaz07}. To do so, we match the stellar mass distribution
of a subsample of the field dwarf catalog to the subsample of
satellite galaxies under study; this ensures that the two samples have
statistically identical distributions of stellar mass. We divide both
samples into bins (typically 10, although we allow this to vary based
on sample size) in the relevant parameter, then randomly select
galaxies with replacement from the sample with the larger number in
that bin, until both subsamples have an equal number in the bin. This
procedure is repeated 100 times, and the results given below are mean
values.

\section{Satellite Quenching as a Function of Central Galaxy Properties}
\label{sec:results}

The primary goal of this paper is to explore the degree to which satellite galaxies
have suppressed or quenched star formation relative to similar galaxies in the field.
In order to numerically interpret our results, we introduce a
conversion fraction (${f^{\rm X}_{\rm convert}}$), designed to indicate the
fraction of galaxies that are converted from star forming to suppressed/quenched
{\em because} they are satellites.  
% More specifically, the conversion fraction
% is the difference in the quenched fraction between the satellite and control
%samples relative to the unquenched fraction of the control sample.

Mathematically, the conversion fraction is defined as follows.  Let 
${\rm X} = \log(\rm SSFR)$ indicate a variable value for the specific star formation 
rate.  Define the unquenched fraction (with $\log{\rm SSFR} \ge X$) of
satellites and control galaxies to be $u_{\rm sat}$ and $u_{\rm control}$, respectively.
The associated quenched fractions with $\log{\rm SSFR} < {\rm X}$ are 
$q_{\rm \{sat,control\}} =1-u_{\rm \{sat,control\}}$.
  The conversion
fraction, $f^{\rm X}_{\rm convert}$, is then given by
\begin{eqnarray}
  \label{eq:conv}
f^{\rm X}_{\rm convert}
&=&\frac{q_{\rm sat}-q_{\rm control}}{u_{\rm control}}\;\bigg |_{{\rm
    SSFR}=10^{\rm X}}\,.
\end{eqnarray}
For example, if 100\% of the galaxies in the control sample are
star forming (at the given SSFR threshold) but only 80\% of the
satellite galaxies are star forming (at the same threshold), then
$f_{\rm convert}=0.2$ and we may conceivably argue that $\sim 20 \%$ of star forming galaxies were 
converted to quenched galaxies after becoming satellites.  Likewise, for star forming fractions of 60\%
and 40\% for the control and satellite samples (respectively), the
conversion fraction would be $f_{\rm convert}=0.33$.

%%%%%%%%%%%%%%%%%%%%%%%%%%%%%%%%%%%%%%%%%%%%%%%%%%%%%%%%%%%%%%
\begin{figure*}
 \centering
 \includegraphics[scale=0.45]{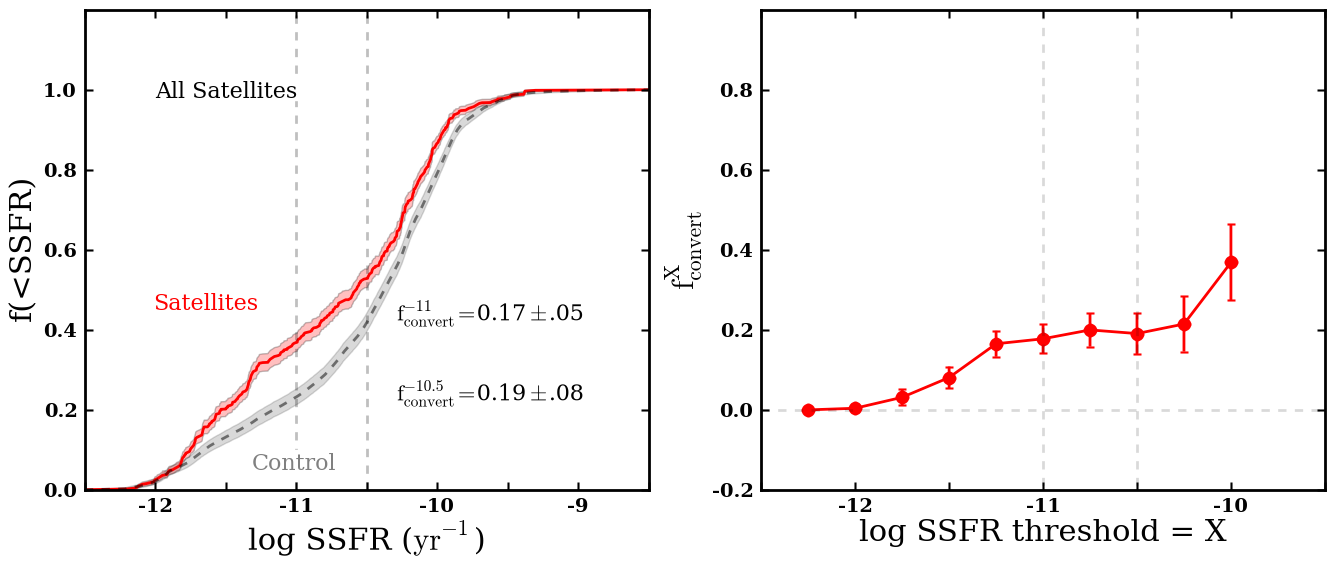}
 \caption{(\emph{Left}): Cumulative distributions of specific star
   formation rate for the observed satellite sample (red dashed
   line), the purity-corrected satellite sample (red solid line), and
   the control sample (black dashed line). After a modest statistical
   correction for interloper contamination, we find that $\lstar$
   hosts quench $\sim \! 20\%$ of their infalling star forming
   satellites. Black and red shaded regions correspond to $1\sigma$
   binomial errors for the control and purity-corrected satellite
   samples, respectively. (\emph{Right}): Conversion
   fraction as a function of specific star formation threshold $X$, which provides
   an estimate for the fraction of satellites that have had their
   star formation suppressed below a threshold value $X$ as a result of becoming satellites.  This
   quantity is defined in equation \ref{eq:conv}. The conversion fraction, $\rm f_{convert}$, is quite flat over a wide range in satellite star formation rates. Vertical dashed lines correspond to our definitions of ``quenched" and ``vigorously star forming" satellites.
 \label{fig:all}
}
\end{figure*}

%%%%%%%%%%%%%%%%%%%%%%%%%%%%%%%%%%%%%%%%%%%%%%%%%%%%%%%%%%%%%%

This definition of conversion fraction is similar to the similar  quenching efficiency specified by e.g. \citet[]{vandenBosch08,peng12}, although
$f^{\rm X}_{\rm convert}$ is generalized to be a function of SSFR.
The two conversion fractions we will primarily consider are ${\rm X} =
- 11$ and ${\rm X} = -10.5$. The conversion fraction evaluated at the
passive SSFR threshold for satellites, $f^{-11}_{\rm convert}$, can be
thought of as the fraction of satellites that were star forming prior
to infall and have subsequently been converted to passive satellite
systems, making it an effective method of quantifying quenching. The
conversion fraction evaluated at the border between vigorous and
moderate star formation, $f^{-10.5}_{\rm convert}$, is representative
of the fraction of vigorous star forming galaxies that upon infall had
their star formation either reduced to a moderate level or halted
entirely. In general, we will refer to a satellite's SSFR being lowered below $10^{-11} \, yr^{-1}$ as ``quenching," and the general case of a satellite's SSFR being lowered as ``suppression" (e.g. we will often discuss $f^{-10.5}_{\rm convert}$ as suppression of vigorous star forming satellites). We will make use of these parameters throughout the
remainder of this work. 

\subsection{All Satellites}

Figure \ref{fig:all} shows the cumulative distributions of SSFRs for
all satellite galaxies in our sample. The dashed red line is the
observed distribution SSFR for the satellite population, while the
solid red line is the corrected satellite distribution, where the
observed distribution is adjusted to take into account the presence of
interlopers. This is done by subtracting the control distribution
scaled by the probability that a randomly selected galaxy is an
interloper, (1-$f_{\rm purity}$), from the satellite distribution. The
adjusted distribution is then re-normalized.\footnote{This assumes
  that the interlopers follow the same specific star formation
  distribution as the control sample.} This correction for purity is
made in each plot of satellite SSFR that follows, and the uncorrected
result is omitted. The black line shows the distribution for the mass-matched field
control sample --- i.e.~for a sample of isolated (central) galaxies
with the same stellar mass distribution as the satellites. For both the corrected satellite and control samples, binomial errors are shown
by the corresponding shaded region according to the formula
\begin{eqnarray}
  \label{eq:conv}
\sigma ^2
&=&\frac{(q_{\rm sample})(u_{\rm sample})}{N_{\rm sample}}
\end{eqnarray}
where $q_{\rm sample}$ is the sample's quenched fraction, $u_{\rm sample}$ is the sample's unquenched fraction, and $N_{\rm sample}$ is the total number of objects in the sample.

Globally, $f^{-11}_{\rm convert} = 0.18$ for our $\lstar$ host sample, with a similar value for
$f^{-10.5}_{\rm convert}$.
This implies that approximately 20\% of
satellites are quenched as a result of falling into their host's virial radius. 
The right panel of Figure \ref{fig:all} generalizes $f_{\rm convert}$ to an arbitrary
SSFR threshold.  Bright satellite galaxies residing in haloes similar to that of the
Milky Way show an overall flat increase in conversion
fraction over a dex in satellite SSFR ranges. We use
this style of presentation in the next three plots; the left panels
show the cumulative distributions of SSFR for the satellite and
control samples and the rightmost panel shows $f_{\rm convert}$
plotted as a function of satellite threshold SSFR (as defined by
Equation \ref{eq:conv}).

%%%%%%%%%%%%%%%%%%%%%%%%%%%%%%%%%%%%%%%%%%%%%%%%%%%%%%%%%%%%%% 

\subsection{Dependence on Host Star Formation Rate}

In order to investigate the correlation of quenching efficiency with
host star formation activity, we define a mass-dependent passive
threshold according to SSFR in our host sample, given as Equation 1
above. Hosts that fall below the relation are considered passive ($N =
267,\, \langle \mstar \rangle = 10^{10.81}~\msun$), while those above it are
considered star forming ($N = 190,\, \langle \mstar \rangle =
10^{10.78}~\msun$).  Our goal of determining the correlation between
star formation activity in satellites and hosts is complicated by the
fact that
% We wished to isolate the correlation with host star formation; to
passive hosts tend to have slightly higher stellar masses. To avoid
biasing our results, we mass-match the host samples --- i.e.~we ensure
that the distribution of stellar masses is the same between our
star forming and passive host samples.

\begin{figure*}
 \begin{centering}
 \includegraphics[scale=0.42, viewport= 0 0 2000 410]{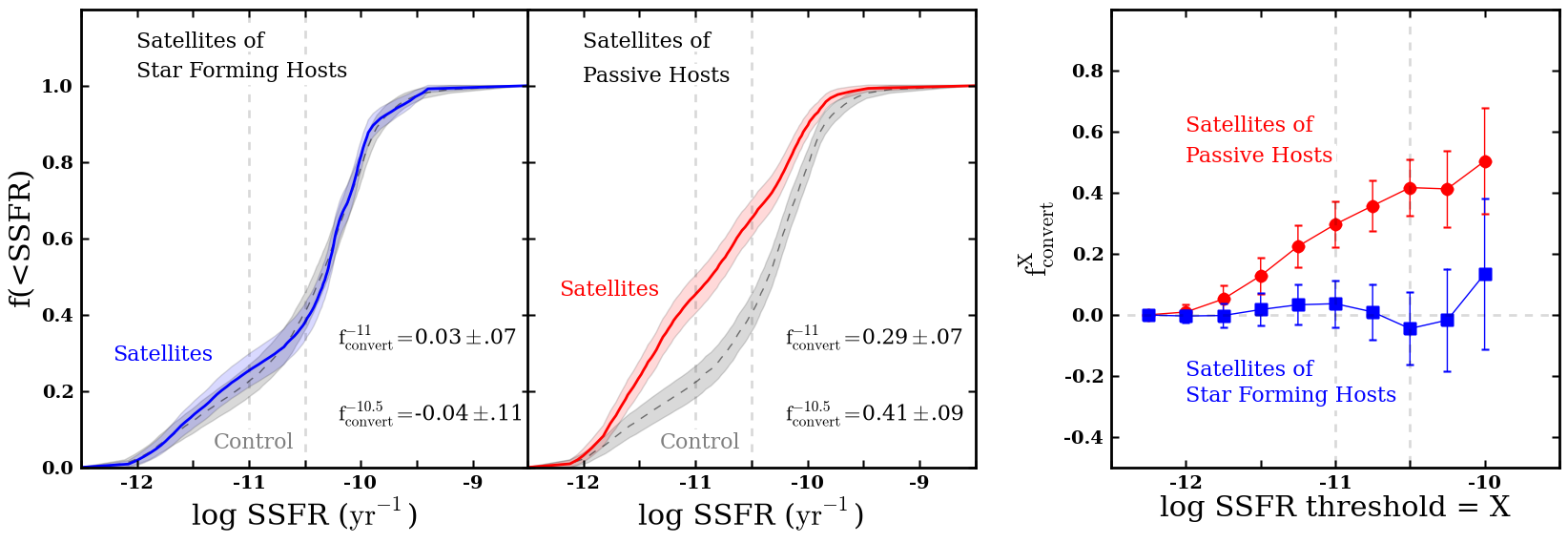}
 \end{centering}
 \caption{(\emph{Left}): Cumulative distributions of specific star
   formation rate for satellites surrounding star forming hosts (blue
   line) and passive hosts (red line), with relevant mass-matched
   control samples (black dashed lines). Passive hosts quench $\sim
   30\%$ of their infalling star-forming satellites, while
   star forming hosts have satellite populations that very closely
   match the field. Shaded regions are binomial errors. (\emph{Right}):
   Conversion fraction vs.~threshold SSFR (see equation \ref{eq:conv})
   for satellites of star forming and passive hosts. At all values of
   $\rm X$, the satellites of star forming hosts are indistinguishable
   from the field population at the same stellar mass.  In stark contrast, satellites of
   passive hosts show an increasing difference from the field at higher
   values of SSFR.
 \label{fig:colour}
}

\end{figure*}

Comparisons of the cumulative distributions of satellite SSFRs for the
two mass-matched host samples are shown in Figure \ref{fig:colour}. The
satellite population around quenched hosts (middle panel) is itself
quenched relative to the mass-matched field sample, with $f^{-11}_{\rm convert}=0.28$ and $f^{-10.5}_{\rm convert}=0.35$.   This result is qualitatively consistent 
with the observed correlation between host and satellite properties
otherwise known as ``galactic conformity'' \citep[e.g.][]{ weinmann06, kauffmann13,
  robotham13}. 
  
Remarkably, the satellite population around
star forming hosts (left panel) is markedly different, with
a SSFR distribution that is \textit{indistinguishable}
from the mass-matched sample of field galaxies. The right panel of
Figure~\ref{fig:colour} emphasizes this stark difference --- at
all choices of SSFR threshold, the conversion fraction for satellites
of star forming hosts (blue line) is consistent with zero. Meanwhile,
the conversion fraction for passive hosts increases monotonically
as the SSFR threshold decreases, suggesting that passive hosts
are more likely to suppress vigorous star formation in their satellites
than strongly quench it to passivity.

Figure~\ref{fig:colour} provides a compelling demonstration that satellite
quenching and central quenching are strongly related: among 
bright $\sim 0.1 \lstar$ satellites, only those that
inhabit quenched hosts are themselves quenched (relative to the field
population). This is somewhat distinct from the view usually 
discussed as galactic
conformity, in which the colour or star formation
rate of a satellite is correlated with the colour or star formation
rate of its host. Rather, it appears that bright satellites of isolated,
star forming $\lstar$ galaxies are essentially unaffected by their
host, whereas approximately 25-35\% of satellites of quenched hosts
are quenched as a result of being a satellite. The implications of
this finding will be discussed in
Section~\ref{sec:discuss}.

%%%%%%%%%%%%%%%%%%%%%%%%%%%%%%%%%%%%%%%%%%%%%%%%%%%%%%%%%%%%%%
%%%%%%%%%%%%%%%%%%%%%%%%%%%%%%%%%%%%%%%%%%%%%%%%%%%%%%%%%%%%%% 

\subsection{Dependence on Satellite Morphology}
\label{subsec:morph}

The previous subsection demonstrated that the quenching of $\sim 0.1 \lstar$ satellites only
occurs around galaxy-size hosts that are themselves quenched. A natural extension
of this result is to ask whether the morphologies of the satellites
are affected along with their star formation rates, where in this work we will use S\'{e}rsic index as a proxy for morphology. In other words, is
the quenching mechanism connected to a morphological change?

\begin{figure*}
 \centering
 \includegraphics[scale=0.44]{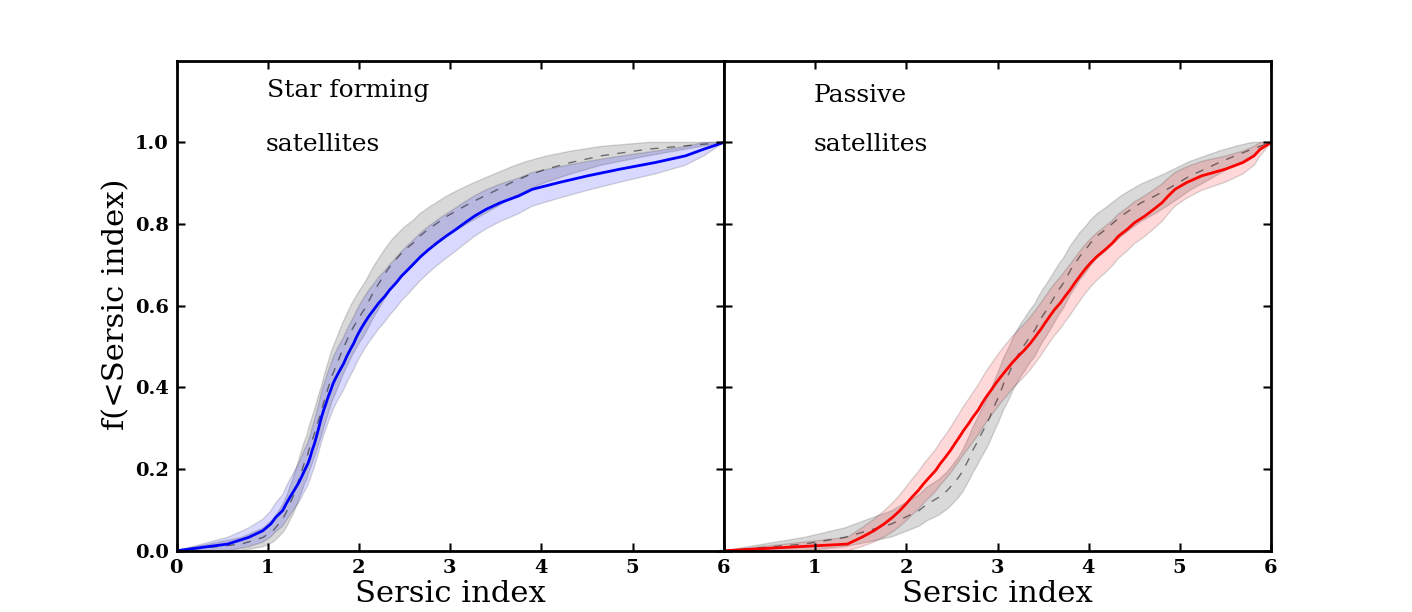}
 \caption{Cumulative distributions of S\'{e}rsic indexes for star forming (left
   panel) and passive (right panel) satellites around passive
     hosts and mass matched control samples
   (black dashed lines). Satellites show no significant 
   difference in morphology relative to field galaxies at fixed stellar mass and
   star formation activity. Shaded regions are binomial errors.
 \label{fig:sersic}
}
\end{figure*}

Figure~\ref{fig:sersic} explores this question by comparing the S\'{e}rsic index distributions for
passive and star forming satellites of quenched hosts to mass-matched
control samples of the same star formation category (passive or star forming).  Neither satellite sample shows substantial deviation from the field. One might
expect to observe no morphological difference between the star forming satellite sample. However, it
is less obvious why the passive field dwarf sample resembles the
passive satellite sample when the field galaxies are manifestly not
being quenched by an environmental process, and a substantial fraction of the passive satellites of passive galaxies are.

It is clear from
Figure~\ref{fig:sersic} that whatever is causing satellite galaxies to be
quenched at a higher rate than field galaxies, this process results in
morphological properties that are indistinguishable from field galaxies
that are presumably quenched via some distinct secular process.
One potential implication is that the morphological differences that divide
quenched and star forming galaxies are a {\em result}
of a galaxy being quenched.  This is distinct from the idea of ``morphological quenching"
that suggests that morphological changes themselves give rise to 
suppressed star formation \citep{martig09}.   We return to a discussion of these issues in Section 5.

\subsection{Dependence on Projected Separation}
\label{subsec:radial}

\begin{figure*}
 \centering
%, viewport=20 0 400 420
 \includegraphics[scale=0.44]{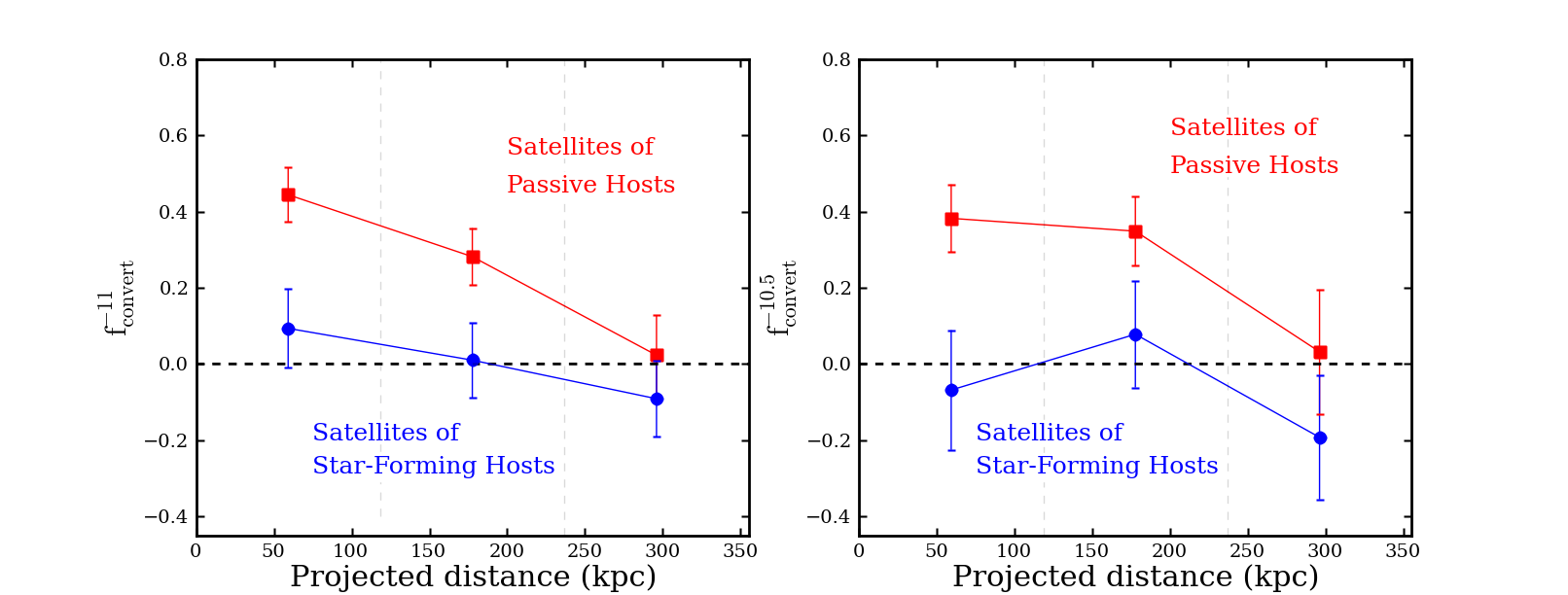}
 \caption{Conversion fractions of satellites around passive (red line)
   and star forming hosts (blue line) as a function of projected
   distance from the central galaxy. The dashed vertical grey lines
   denote the edges of the radial bins. (\emph{Left}): $f^{-11}_{\rm
     convert}$, representative of the fraction of star-forming
   satellites that have their star formation quenched upon
   infall. Around star-forming hosts, quenching is only seen at small
   projected radii, while around passive hosts, quenching is observed
   out to our radial limit. (\emph{Right}): $f^{-10.5}_{\rm convert}$,
   representative of the fraction of vigorously star forming
   satellites that have their star formation suppressed upon
   infall. For star forming hosts, the quenching observed in the left
   panel is no longer apparent; the frequency of vigorous star forming
   satellites suppressed around star forming hosts is consistent with
   the field at all projected radii.
  \label{fig:radsfr}
}
\end{figure*}

We expect that the efficiency of some potential quenching processes
may vary with the separation of the host and satellite,
possibly reflecting radial variation in the circumgalactic medium
(CGM) surrounding the host, the strength of the local tidal field, or
the time since first infall for the satellite. In order to investigate
trends with host/satellite distance, we place our satellite sample into
three linearly spaced bins in projected separation. Conversion
fractions are shown as a function of projected distance from the host,
with samples divided according to SSFR in Figure
\ref{fig:radsfr}. Subsamples in each bin of projected separation are
matched on host mass and compared to control samples matched to the
corresponding satellite mass distributions. When examining the
dependence of $f^{11}_{\rm convert}$ on satellite/host separation, we
find a gradient in quenching efficiency.  Around
passive hosts, quenching persists only out to the intermediate
projected radius bin, which has its outer edge at 236 kpc, less than
the characteristic virial radius of our primary sample in MS-II ($\sim
\! 400~\kpc$).
Around star forming hosts, however, we find results consistent with no
quenching at all radii. 

The right panel of Figure
\ref{fig:radsfr} shows $f^{10.5}_{\rm convert}$, which is sensitive to the {\em damping} of vigorous star
formation, if not full quenching.  The trends are highly similar in shape; the largest
difference is the central value in the innermost bin of the
star forming host sample.  Again, we see no evidence of quenching
around star forming hosts, and strong quenching around passive hosts.

%%%%%%%%%%%%%%%%%%%%%%%%%%%%%%%%%%%%%%%%%%%%%%%%%%%%%%%%%%%%%%
%%%%%%%%%%%%%%%%%%%%%%%%%%%%%%%%%%%%%%%%%%%%%%%%%%%%%%%%%%%%%% 

\section{Discussion}
\label{sec:discuss}

The main result from our study of central and satellite galaxies in
the SDSS is that bright satellites of $\sim \!  L^{*}$ galaxies are
only quenched beyond what is seen in the field around passive hosts, indicating a strong connection
between the quenching of satellites and centrals in Milky Way-like
haloes.  This host-specific dichotomy in satellite quenching is likely (or at least partially)
responsible for the observed correlation between the properties of
central and satellite galaxies in the local Universe, a phenomenon
commonly dubbed ``galactic conformity.'' As first shown by
\citet{weinmann06}, studying more massive haloes than those isolated in
our work, groups dominated by a ``late-type'' central host a higher
fraction of ``late-type'' satellites relative to groups populated by
an ``early-type'' central.\footnote{Note that the terms late- and
  early-type as used by \citet{weinmann06} correspond to the colour
  and/or SSFR of the galaxy and do not refer to a morphological
  classification.} More recently, results from \citet{wang12}
highlight a similar conformity in the rest-frame $g-r$ colour of
satellite galaxies in the SDSS, examining systems across a broad range
of host stellar
mass. 
Lastly, studying low-mass galaxy groups identified by the Galaxy And
Mass Assembly project \citep[GAMA,][]{driver11}, \citet{robotham13}
find that conformity is also evident in the morphological properties
of satellites relative to their hosts, such that bulge-dominated
central galaxies tend to host bulge-dominated satellites in contrast
to disk-dominated centrals. Previous works have demonstrated that semi-ananlytic models do show evidence of galactic conformity, however the nature of the conformity does not match observations. For example, \citet{wang12} argue that the semi-analytic model of \citet{guo11} overpredicts the number of red satellites in g-r (colour). Similarly, we have analyzed the \citet{guo11} model results independently and found that they predict a conversion fraction of ~0.6 around passive hosts and ~0.3 around star-forming hosts, where we apply our same isolation criteria to the SAM. This is in stark disagreement with the dichotomy we see in the data.

In general, our results are in agreement with each of the
aforementioned observational analyses. However, unlike these previous studies that
compare the composition of $\gtrsim 10^{12}~\msun$ haloes separated
according to the properties of the central galaxy (i.e.~excluding the
field population), our analysis extends this work to focus on the
comparison of satellite samples (again divided according to the
properties of the central) to field galaxies of like stellar
mass. This is a critical step towards understanding the physical
driver of satellite quenching in the local Universe, as it directly compares
satellite galaxies to their parent population (i.e.~field systems of
like mass).  

One possible explanation for the observed dichotomy in satellite
quenching evident in our sample could be a dependence of quenching
efficiency on halo mass. That is, passive hosts could live in more
massive dark matter haloes relative to their star forming counterparts
at fixed stellar mass, with quenching operating more efficiently in
more massive haloes. As detailed in Section \ref{sec:iso_sdss}, we
tailored our host selection criteria to isolate a sample that spans a
relatively small spread in halo mass (FWHW $\sim 0.5$ dex) and peaks
at a~few~$\times 10^{12}~\msun$.  Studies of
gas accretion onto dark matter haloes indicate that there is a
transition in the dominant accretion mode at roughly this halo mass,
with infalling gas shock-heated at the virial radius in haloes above a threshold mass of
$\sim$~a~few~$\times 10^{12}~\msun$, while cold gas reaches a smaller radius, possibly
falling all the way to the galaxy in
 less-massive haloes \citep[e.g.][]{binney77, rees77,
  birnboim03, keres05, keres09,Stewart11a}.   Such a variation in accretion mode could manifest itself as a
correlation between the star forming properties of central (and also
satellite) galaxies and the mass of their host dark matter haloes, such
that passive central galaxies reside in more massive dark matter haloes
relative to star forming central galaxies of comparable stellar
mass.    

Recall that in our analysis, we tune our isolation criteria to
identify haloes comparable in mass to that of the Milky Way, but select
our sample of star forming and passive hosts for comparison by matching them
in stellar mass (i.e.~potentially allowing a weak
correlation between halo mass and host properties). In Figure
\ref{fig:satellite_velocities}, we investigate this possibility by
plotting separately the stacked velocity distribution of stellar
mass-matched subsamples of satellites around our passive and
star forming host galaxies. While we find that the red hosts are
biased towards slightly more massive haloes (approximately $45\%$ more massive), the difference is only
significant at the $\sim \! 1-2\sigma$ level.
 Furthermore, previous work
examining galactic conformity by \citet{weinmann06} found that the
correlation between the colour of central and satellite galaxies exists
at fixed halo mass, looking at groups in the \citet{yang07}
catalog, where the halo masses computed in the group catalog are based on the total luminosity or stellar mass of all member galaxies above a given luminosity
threshold. This suggests that a potential dependence of quenching
efficiency on halo mass is unable to explain our results in concert
with those of \citet{weinmann06}; while we find evidence for
conformity at fixed {\em stellar} mass (and similar, though possibly not identical halo mass),
\citet{weinmann06} claims conformity at fixed {\em halo} mass, although that result is consistent with ours.

The dichotomy in satellite quenching also could be driven by
differences in the circumgalactic medium (CGM) of passive and
star forming hosts.
The CGM of the host halo could potentially affect
the star formation of a satellite galaxy via two physical processes:
(\emph{i}) ``ram pressure stripping,'' where the cold interstellar gas
from which stars could be actively formed is removed from the
satellite, rapidly truncating its star formation
\citep[e.g.][]{gunn72, bekki09} and (\emph{ii}) ``strangulation,''
where the satellite retains its cold gas, allowing active star
formation, but loses its reservoir of warm gas that would otherwise
replenish the cold gas as it is exhausted \citep[e.g.][]{larson80,
  kawata08}. In either scenario, our results could be explained by the
presence of a hot gaseous corona preferentially surrounding passive
hosts, such that it prevents the infall of cold gas onto the central
galaxy, thereby quenching it, while similarly halting the star
formation of associated satellites.   A CGM dichotomy could
also arise in association with the the transition between cold and hot mode accretion.
While realistically we expect there to be some scatter in the halo mass scale where the transition
occurs  \citep{keres09}, there might naturally be a correlation between star formation in the
central galaxy and ongoing cold-mode accretion.  In hot mode haloes, star formation may
be naturally suppressed by a lack of fuel, and the associated build-up of
a hot corona could then provide a CGM-related suppression mechanism.

In our discussion of the hot CGM interpretation, we must adress the results of \citet{tumlinson11}: quasar line-of-sight
probes of the CGM of nearby systems show an enhancement in ionized oxygen
around star forming galaxies relative to their quenched counterparts, indicating that star forming galaxies host more
warm gas in their haloes. The lack of O {\scriptsize VI}
around passive hosts could be explained as a temperature effect, with passive
galaxies surrounded by a typically hotter CGM. Observations of X-ray
coronae around local ellipticals support this conclusion
\citep{osullivan01, sun07}, including recent work to study the hot gas
surrounding lower-mass, isolated ellipticals \citep[][]{mulchaey10,humphrey11,humphrey12}.

%%%%%%%%%%%%%%%%%%%%%%%%%%%%%%%%%%%%%%%%%%%%%%%%%%%%%%%%%%%%%%
%%%%%%%%%%%%%%%%%%%%%%%%%%%%%%%%%%%%%%%%%%%%%%%%%%%%%%%%%%%%%% 

In Section \ref{subsec:morph}, we also examined the morphology of the
satellites in our sample, finding consistent S\'{e}rsic indices at fixed
stellar mass between the quenched satellites and
quenched galaxies in the field. The same is true for star forming satellites. This result could be compared to similar results in \citet{vandenBosch08}, who found consistent concentrations between satellites and centrals at fixed stellar mass and colour, and \citet{bamford09} who found no trend in the early-type fraction of galaxies with environment in the sparse regime (and a weak trend, as compared to the trend in red fraction, over all environments).
The lack of any difference between satellite and field morphology could
provide a clue to the origin of the well-established link between structure and star
formation in galaxies \citep[e.g.][]{kauffmann03, blanton05,wuyts11,bell12,Cheung12},
with quenched galaxies typically having more bulge-dominated
morphologies than star forming ones. 
  
One idea to explain the correlation between morphology and star formation activity posits that the morphological transformation itself is the driving agent.   Specifically, in the  ``morphological quenching'' scenario \citep{martig09}, the presence of a bulge component stabilizes a galaxy
against fragmentation to bound clumps, thereby halting star
formation. In such a picture, a bulge-dominated galaxy may remain
quenched even if it accretes gas (i.e.~regardless of the assumed halo
mass-dependent accretion mode).  Our results may be at odds with such a scenario: satellites that are apparently quenched via environmental mechanisms never experienced by quenched galaxies in the field nevertheless have exactly the same distribution of morphological indexes.  If morphology  were really the driver of quenching, it seems unlikely that this would be the case, unless somehow the passive-halo environment causes exactly the same type of bulge formation as do processes in the field.  One might argue that tidal forces could incite secular instabilities, but this would also produce quenching for satellites around non-star forming hosts.  No such signal is seen.

  Often, galaxy mergers are relied upon as a means for
driving the correlation between star formation activity and morphology
\citep{springel05, hopkins06}. However, our results suggest that
mergers, which occur preferentially in group-like environments or the field
\citep{mcintosh08, fakhouri08, darg10, lin10}, are unlikely to be
responsible for the properties of quenched satellites in passive 
galaxy haloes. Instead, whatever mechanism(s) is responsible for
shaping the morphology of our quenched satellite population needs to
operate as effectively in the field (i.e.~lower-mass haloes) as it does
for subhaloes within Milky Way-like systems. While many models suggest that bulge
formation is merger-driven \citep[e.g.][]{toomre77, springel05,
  cox06}, our results would favor a different origin for bulge-dominated
morphology \citep[e.g.][]{kormendy04, jogee05, macarthur03}, since we
find no variation in S\'{e}rsic index with environment.
Collectively, our results may support a picture where the cessation of star formation (e.g.~via gas removal from
ram pressure/strangulation or blow-out) triggers  
an associated morphological transformation.

Finally, we also examined the dependence of quenching efficiency on
projected host/satellite separation, finding a radial gradient in the
conversion fraction, $f^{\rm X}_{\rm convert}$, around passive
galaxies. Recent work by \citet{kauffmann13} reports
a correlation in the specific SFR of satellites extending out to
comparable spatial scales around massive central galaxies; for hosts
with stellar masses of $> \!  10^{10}~\msun$, \citet{kauffmann13}
argue that this correlation is confined to spatial scales less than $1$-$2$
Mpc. Analyzing satellite galaxies around massive ($\sim \!
10^{11}~\msun$) central galaxies selected from the GAMA project,
\citet{prescott11} also find a significant radial dependence for the
satellite red fraction out to separations of roughly $500$ Mpc.

The observed radial trend measured in our sample could reflect a
correlation between infall time and the cessation of star formation,
such that more-recently accreted satellites are more likely to be
star forming \citep[see also][]{hearin13}. However, the connection between infall time and
projected distance is particularly poor inside the virial radius
\citep{oman13}. Instead, if the CGM drives the quenching of star
formation via ram-pressure stripping or strangulation, the radial
dependence of quenching efficiency could be the result of radial
variation in the properties of the CGM --- a denser CGM would yield
more effective gas stripping, as the ram pressure force scales with
the density of the stripping medium \cite{gunn72}.

Altogether, our study has uncovered two remarkable results with regard
to the properties of relatively massive satellite galaxies around
$\sim \! L^{*}$ hosts in the local Universe: (\emph{i}) falling into a
$\sim \! 10^{12.3}\msun$ halo hosted by a star forming central
galaxy has a negligible impact on the star formation rate of
$\sim~\!~0.1$ $L^{*}$ galaxies, such that massive satellites of
star forming hosts are indistinguishable from a field population of
like stellar mass, and (\emph{ii}) $\sim \! 30\%$ of star forming
galaxies accreted into a similar halo hosting a passive central galaxy
are quenched, but with no corresponding change in morphology. When
combined, these results support a picture where the mechanism(s) by which
bright satellites of $\sim \!  L^{*}$ galaxies are quenched only occur in haloes 
where the central galaxy is itself quenched.  This could either imply
that the same quenching mechanism is operating on host and satellite, or
it could point to a secondary effect, perhaps associated with differences in the CGM
of quenched and star forming galaxies.
Whatever the quenching mechanism, it impacts the morphologies of 
satellites in a manner that mirrors distinct quenching mechanisms in the field.

It is worth emphasizing one particularly interesting implication of our findings:
 even for passive hosts, the
quenching mechanism(s) must operate(s) inefficiently \citep[possibly with a long timescale,][]{delucia12, wetzel13a,trinh13}, such that only $\sim \!  30\%$ of 
satellites that have fallen in from the field are quenched. This
relatively low quenching efficiency is mirrored in studies of
lower-mass satellites \citep{geha12}.   It is not clear why quenching occurs around 
passive hosts (and not star forming hosts) but the fact that it is apparently fairly inefficient may
be a clue to its origin.  Moreover, while our results focus on fairly massive satellites
($\mstar > 10^{9.5}~\msun$), we know from studies of dwarf galaxies in the Local Group that
quenching must become very efficient for smaller satellites. 
The vast majority of low-mass satellites  ($\mstar < 10^{8}~\msun$) within the virial radii of the Milky Way and M31
are gas-poor and quenched, in contrast to the field population \citep{mcconnachie12}. Our ongoing
work involves investigating this potential mass dependence along with
the exploration of simple quenching models that can replicate the
relatively inefficient quenching of massive satellites (Wheeler et al., in
prep).

\section{Conclusions}
\label{sec:summary}

In this work, we have investigated quenching and suppression of star
formation of satellite galaxies in the mass range $\rm 10^{9.5} \msun
< \mstar < 10^{10.5} \msun$ orbiting hosts that are selected to have
$\mstar > 10^{10.5}~\msun$.  Our hosts are chosen 
to be central galaxies of  $\sim 10^{12.3} \msun$ dark matter haloes, selected via careful
isolation criteria to the group and cluster environment.  In order to evaluate
the degree of quenching among satellite systems, we make comparisons to 
stellar-mass matched control samples of isolated, low-mass field galaxies.

Our key result  (Figure 7) is that there is a dichotomy in satellite quenching based on host star formation
strength: passive hosts quench roughly 30\% of their infalling
satellites, while satellites of star forming hosts {\em of the same stellar mass} show a SSFR
distribution consistent with field
galaxies. 
Around passive galaxies, quenching is only present in the inner $\sim
200 \,\kpc$, significantly less than the virial radius of these hosts,
and increases in efficiency with decreasing host/satellite
separation. We also show that passive satellites and passive field
galaxies show no morphological distinction at fixed stellar mass.

This paper presents a clear dichotomy in massive satellite
quenching; however, we anticipate that other trends will be
present. For example,  smaller ($\mstar < 10^8 \msun$) satellites in the Local Group 
are universally quenched, suggesting that quenching might be related to satellite
stellar mass. We will explore trends with host and satellite stellar
mass in a companion paper.

\section*{Acknowledgments} 
This work was supported in part by NSF grants AST-1009973 and AST-10999 and
NASA grant NNX09AD09G.  MB-K acknowledges support from the Southern California Center for
Galaxy Evolution, a multicampus research program funded by the
University of California Office of Research. The Millennium-II
simulation database used in this paper and the web application
providing online access to them were constructed as part of the
activities of the German Astrophysical Virtual Observatory. EJT
acknowledges support for this work provided by NASA through Hubble
Fellowship grant \#HF-51316.01, awarded by the Space Telescope Science
Institute, which is operated by the Association of Universities for
Research in Astronomy, Inc., for NASA, under contract NAS 5-26555.

Funding for the SDSS and SDSS-II has been provided by the Alfred
P. Sloan Foundation, the Participating Institutions, the National
Science Foundation, the U.S. Department of Energy, the National
Aeronautics and Space Administration, the Japanese Monbukagakusho, the
Max Planck Society, and the Higher Education Funding Council for
England. The SDSS Web Site is http://www.sdss.org/.

The SDSS is managed by the Astrophysical Research Consortium for the
Participating Institutions. The Participating Institutions are the
American Museum of Natural History, Astrophysical Institute Potsdam,
University of Basel, University of Cambridge, Case Western Reserve
University, University of Chicago, Drexel University, Fermilab, the
Institute for Advanced Study, the Japan Participation Group, Johns
Hopkins University, the Joint Institute for Nuclear Astrophysics, the
Kavli Institute for Particle Astrophysics and Cosmology, the Korean
Scientist Group, the Chinese Academy of Sciences (LAMOST), Los Alamos
National Laboratory, the Max-Planck-Institute for Astronomy (MPIA),
the Max-Planck-Institute for Astrophysics (MPA), New Mexico State
University, Ohio State University, University of Pittsburgh,
University of Portsmouth, Princeton University, the United States
Naval Observatory, and the University of Washington.

The authors wish to thank Betsy Barton for her support and contributions to this work, and wish to thank the anonymous referee for their helpful comments.

\bibliography{dichotomy_draft.bib}

\label{lastpage}
\end{document}